\documentclass[aps,pre,twocolumn,groupedaddress,showpacs,showkeys,preprintnumbers]{revtex4}

\usepackage{graphicx}

\begin{document}

 \preprint{GCEP-CFUM/2006/AC01}

 \title{Gap-Size Distribution Functions of a Random Sequential Adsorption Model of Segments on the Line}

 \author{N.A.M. Ara\'ujo}
 \author{A. Cadilhe}
 \email[Corresponding author: ]{cadilhe@fisica.uminho.pt}
 \affiliation{GCEP-Centro de F\'isica da Universidade do Minho, 4710-057 Braga, Portugal}

 \date{\today}

 \begin{abstract}
  We performed extensive simulations accompanied by a detailed study of a two-segment size random sequential model on the line.
  We followed the kinetics towards the jamming state, but we paid particular attention to the characterization of the jamming state structure.
  In particular, we studied the effect of the size ratio on the mean-gap size, the gap-size dispersion, gap-size skewness, and gap-size kurtosis at the jamming state.
  We also analyzed the above quantities for the four possible segment-to-segment gap types.
  We ranged the values of the size ratio from one to twenty.
  In the limit of a size ratio of one, one recovers the classical car-parking problem.
  We observed that at low size ratios the jamming state is constituted by short streaks of small and large segments, while at high values of the size ratio the jamming state structure is formed by long streaks of small segments separated by a single large segment.
  This view of the jamming state structure as a function of the size ratio is supported by the various measured quantities.
  The present work can help provide insight, for example, on how to minimize the interparticle distance or minimize fluctuations around the mean particle-to-particle distance.
 \end{abstract}


 \maketitle

 \section{Introduction\label{introduction}}
  Since the 1939 paper by Flory \cite{Flory39}, for the deposition of dimers on lattices, and the 1958 paper by R\'enyi \cite{Renyi58,Renyi63,Gonzalez74}, for the deposition of segments on a line, the random sequential adsorption (RSA) model has become a paradigm for the study of many natural phenomena, not only in the traditional area of physical chemistry (reaction in polymer chains, chemisorption, colloids, etc), but also in the less traditional areas such as biology, ecology, and sociology \cite{Evans93, Privman94,Privman97,Marro99,Privman00a,Privman00b}.
  For instance, the use of RSA to determine the limiting coverage on surfaces requires an uniform deposition from stabilized, diluted suspensions of particles sized from 100~\AA\ up to a micron by convective flow \cite{Privman00b}.
  Actual objects involve proteins and submicron colloidal particles \cite{Onoda86}.
  Recently, experimental interest has also included deposition on patterned surfaces, prepared by lithographic methods \cite{Chen02}, and some theoretical work has been also performed \cite{Cadilhe04,Araujo04a}.
  The basic dimer deposition model has suffered several extensions and generalizations, namely, cooperative sequential adsorption with adsorption rates dependent on the local environment \cite{Evans93}, inclusion of relaxational mechanisms such as detachment \cite{Tassel00,Krapivsky94} and diffusion \cite{Bonnier97,Nielaba97}, and multi-layer deposition \cite{Nielaba95,Privman92}.
  Also, the RSA model and its extensions have been studied in one \cite{Fan91a,Fan91b,Bartelt91,Bonnier94,Evans97,Blaisdell70,Privman91,Bartelt94,Dickman91,Evans84a,Privman93,Bonnier01,Hassan01,Hassan02} and two dimensions \cite{Blaisdell70,Evans97,Privman91,Brilliantov96,Dickman91,Evans84a,Evans84b,Swendsen81} either in the continuum and lattice versions.
  More extended accounts can be found on recent reviews by Privman \cite{Privman94,Privman97,Privman00a,Privman00b} and Evans \cite{Evans93}.

  Quite recently, interest in the field has shifted towards the competitive deposition of mixtures of segments with different sizes on the line, with some controversial results \cite{Bonnier01,Hassan02}.
  Apart from issues concerning the competitive deposition and consequent adsorption of particles at interfaces, there is a strong motivation for the study of the resulting "patterned" structure, either induced by the kinetics of deposition \cite{Cadilhe04} or by more controlled means, e.g., by patterning the surface available for deposition \cite{Araujo04a}.
  In this paper, we focus our interest on the study and characterization of the inter-particle distance distribution functions of binary mixtures in one dimension.
  We were, thus, able to perform a more detailed study of subtle correlations developed during deposition, by measuring not only the time dependence of the coverage, but also, more refined quantities, such as the distribution functions of the distance between particles.
  From these basic measurements of particle-to-particle distance at the jamming state, we studied the size ratio dependence of the first four cumulants.
  More specifically, since the third and fourth cumulants are straightforwardly related to the skewness and kurtosis, respectively, we actually used the latter quantities in order to characterize the gap-size distribution functions.
  We observed non-trivial effects, as we varied the size ratio of the segments being deposited, and we were able to explain phenomenologically some of the qualitative features observed in our simulation.
  Therefore, it is justifiable to put some effort to understand such a clean study case, where one can learn the actual effects leading to the cooperative behavior induced by deposition towards the jamming state.
  Moreover, despite our model being one-dimensional, our analysis can serve as a guide to interpret and/or compare with similar results in higher dimensions.
  The paper is organized as follows: we present the model and the particulars of the simulations in Section~\ref{model}.
  In Section~\ref{discussion}, we present our results and discuss them.
  Finally, we present our conclusions in Section~\ref{conclusion}.

 \section{The Model and Theory\label{model}}
  We consider the competitive deposition on the line of segments of two different sizes, namely short segments, which we denote as  $A$-segments, and long ones, which we denote as $B$-segments, under the condition that they must not overlap with each other upon adsorption, and therefore, mimic an excluded volume, short-range interaction.
  The fraction of the line occupied by the adsorbed segments defines the coverage.
  We notice that it is possible to re-scale, without loss of generality, the length scale of the system so that $A$-segments are of unit length, while the size of $B$-segments is $R$.
  In this respect, one can regard the ratio of the length of $B$-segments relatively to the length of $A$-segments as the size ratio,

  \begin{equation}
   R=\frac{\mbox{Length of a $B$-segment}}{\mbox{\ Length of an $A$-segment\ }}\ \ .
  \end{equation}

  \noindent The deposition flux represents the number of incoming segments per unit length (in one dimension) and per unit time.
  Let us denote by $\alpha$ the corresponding deposition flux of $A$-segments and by $\beta$ that of $B$-segments, therefore, having a total incoming flux of segments, $\alpha+\beta$.
  The probability of having an $A$-segment attempting deposition on the line during an interval of time d$t$ is

  \begin{equation}
   p_A=\frac{\alpha}{\alpha+\beta}\ \ ,
  \end{equation}

  \noindent while that of a $B$-segment is

  \begin{equation}
   p_B=\frac{\beta}{\alpha+\beta}\ \ ,
  \end{equation}

  \noindent i.e., $p_B= 1-p_A$ as expected.

  Adsorption on the line can only take place if the incoming particle does not overlap with a previously adsorbed segment, thus mimicking an excluded volume, short-range interaction.
  To make a more straightforward comparison with experimental results, we measure time in terms of the number of layers of segments, which attempted deposition whether these segments actually adsorb on the substrate.

  We performed a series of Monte Carlo simulations for various values of the size ratio and equal fluxes of incoming segments.
  To reduce the uncertainties in the various quantities, it is more relevant to increase the size of the system than the number of samples.
  Therefore, to characterize the jamming state, we simulated a system size of $10^7$ units and generated $10^2$ samples.
  Whenever we followed the time dependence, we used a different algorithm to simulate, which is computationally more demanding both in time and allocated memory.
  Consequently, the simulated system size is smaller, $10^4$, and we let time evolve up to $100$ units, and obtained $10^2$ samples.
  As soon as the total size of the segments, which attempt adsorption on the line, equals the size of the system, we increase time by one unit, regardless of the fact that the segments actually adsorb, or not.

  We now present some useful definitions and relations.
  For the sake of simplicity, the definitions are valid for the jamming state.
  However, extension to intermediate states at specific times only requires the explicit time dependence taken into account and also set the upper limit of integration to infinity to account for gap sizes of all lengths.
  The probability distribution of empty space is defined as

  \begin{widetext}
   \begin{equation}
    P_\emptyset(x)\mbox{d}x= \frac{\mbox{\ Number of empty intervals of size $x$ within d$x$\ }}{\mbox{Total number of empty intervals}}\mbox{d}x\ \ ,
   \end{equation}
  \end{widetext}

  \noindent in the limit of $\mbox{d}x$ being an infinitesimal quantity and the number of ensembles going to infinity.
  Thus, $P_\emptyset(x)\mbox{d}x$ for the jamming state has the property

  \begin{equation}
   \label{normalization}
   \int_0^1 P_\emptyset(x)\mbox{d}x=1 \ \ .
  \end{equation}

  \noindent Therefore, the jamming coverage is given by,

  \begin{equation}
   \theta_J=1-\int_0^1 xP_\emptyset(x)\mbox{d}x=1-<x>_\emptyset \ \ .
  \end{equation}

  Discriminating all gaps between pairs of consecutively adsorbed segments by $AA$, $AB$, $BA$, and $BB$ and defining density distribution functions $P_b(x)$, with $b\in \{AA, AB, BA, BB\}$, one obtains the relation

  \begin{equation}
   \label{discrimination}
   P_\emptyset(x)=P_{AA}(x)+2P_{AB}(x)+P_{BB}(x) \ \ ,
  \end{equation}

  \noindent where we exploited the fact $P_{AB}(x)=P_{BA}(x)$ for the present random sequential adsorption model.
  Note from equations (\ref{normalization}) and (\ref{discrimination}) that the $P_b(x)$ are not normalized.
  Keeping in mind the above definitions, it is now straightforward to reckon higher moments of the gap-size distribution functions, defined by

  \begin{equation}
   \label{moments}
   <x^n>_a=\frac{\int_0^1x^nP_a(x)\mbox{d}x}{\int_0^1P_a(x)\mbox{d}x} \ \ ,
  \end{equation}

  \noindent with $a\in \{\emptyset, AA, AB, BA, BB\}$.
  Using equation (\ref{discrimination}) one can relate the moments of the gap distribution functions with the corresponding moment of the global distribution function given by equation (\ref{moments}) yielding

  \begin{eqnarray}
   <x^n>_\emptyset&=&<x^n>_{AA}\int_0^1P_{AA}(x)\mbox{d}x\nonumber \\
   &&\ \ \ +2<x^n>_{AB}\int_0^1P_{AB}(x)\mbox{d}x\nonumber \\
   &&\ \ \ +<x^n>_{BB}\int_0^1P_{BB}(x)\mbox{d}x \ \ ,
  \end{eqnarray}

  \noindent defined for all values of $n=$ $0$, $1$, $2$, $\dots$
  We also compute the cumulants, $\kappa_m^a$, of a distribution function defined as,

  \begin{equation}
   \label{cumulants}
   \ln G_a(k)=\sum_{m=1}^\infty\frac{(ik)^m}{m!}\kappa_m^a \ \ .
  \end{equation}

  \noindent where $G_a(k)$ is the so called characteristic function defined by \cite{Kampen92}

  \begin{equation}
   \label{characteristic}
   G_a(k)=<e^{ikx}>_a=\frac{\int_0^1e^{ikx}P_a(x)\mbox{d}x}{\int_0^1P_a(x)\mbox{d}x} \ \ .
  \end{equation}

  \noindent Therefore, from equations (\ref{moments}), (\ref{cumulants}), and (\ref{characteristic}), one derives the first four cumulants as

  \begin{eqnarray}
   \kappa_1^a&=&<x>_a\ \ ,\label{cumulant_k1}\\
   \kappa_2^a&=&<x^2>_a-<x>_a^2\ \ ,\\
   \kappa_3^a&=&<x^3>_a-3<x^2>_a<x>_a\nonumber\\
   &&\ \ \ \ \ \ +2<x>_a^3\ \ ,\\
   \kappa_4^a&=&<x^4>_a-4<x^3>_a<x>_a\nonumber\\
   &&\ \ \ \ \ \ -3<x^2>_a^2+12<x^2>_a<x>_a^2\nonumber\\
   &&\ \ \ \ \ \ \ \ \ \ \ \ -6<x>_a^4\ \ ,\label{cumulant_k4}
  \end{eqnarray}

  \noindent where $\kappa_1$� is just the mean value and $\kappa_2$� is the variance.
  The third and fourth cumulants are used in the definition of the skewness,

  \begin{equation}
   S_a=\frac{\kappa_3^a}{\left(\kappa_2^a\right)^{3/2}} \ \ ,\label{skewness}
  \end{equation}

  \noindent and kurtosis,

  \begin{equation}
   K_a=\frac{\kappa_4^a}{\left(\kappa_2^a\right)^2} \ \ ,\label{kurtosis}
  \end{equation}

  \noindent respectively \cite{Kampen92}.

  \begin{table}
   \caption{
    The table presents the coverage, $\theta_J$, as a function of the size ratio, $R$, while $\sigma$ represents the associated error.
    These results are for a system size of $10^7$ and for $10^2$ samples.
    \label{table:jc}
   }
   \begin{ruledtabular}
    \begin{tabular}{c c c c c}
     &$R$&$\theta_J$&$\sigma$&\\
     \hline\\
     &1.00&0.7475958&0.000067&\\
     &1.05&0.7544753&0.000062&\\
     &1.10&0.7599829&0.000063&\\
     &1.20&0.7683652&0.000058&\\
     &1.25&0.7716358&0.000063&\\
     &1.30&0.7744377&0.000054&\\
     &1.40&0.7789946&0.000063&\\
     &1.50&0.7825520&0.000064&\\
     &1.60&0.7854349&0.000053&\\
     &1.75&0.7890016&0.000066&\\
     &2.00&0.7941038&0.000058&\\
     &2.10&0.7961414&0.000053&\\
     &2.25&0.7991661&0.000058&\\
     &2.50&0.8036444&0.000060&\\
     &2.75&0.8073996&0.000065&\\
     &3.00&0.8105172&0.000066&\\
     &3.25&0.8131850&0.000060&\\
     &3.50&0.8155590&0.000065&\\
     &3.75&0.8176675&0.000073&\\
     &4.00&0.8195500&0.000070&\\
     &4.50&0.8227540&0.000063&\\
     &5.00&0.8253667&0.000067&\\
     &5.50&0.8275848&0.000074&\\
     &6.00&0.8294481&0.000076&\\
     &7.00&0.8324537&0.000076&\\
     &8.00&0.8347729&0.000094&\\
     &10.00&0.8381112&0.000097&\\
     &12.00&0.8403695&0.00010&\\
     &14.00&0.8420603&0.00011&\\
     &16.00&0.8433187&0.00013&\\
     &18.00&0.8443186&0.00012&\\
     &20.00&0.8451377&0.00012&\\
    \end{tabular}
   \end{ruledtabular}
  \end{table}

 \section{Results and Discussion\label{discussion}}

\begin{figure}
   \includegraphics[width=8.5cm]{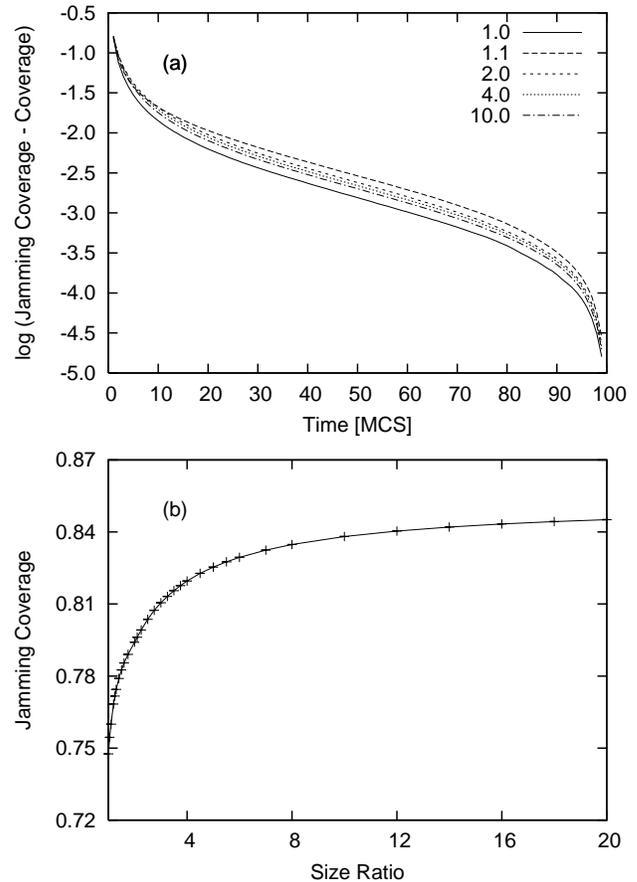}
   \caption{
    Coverage dependence on the size ratio for equal depositing fluxes of each segment size.
    $\mbox{a)}$ Plot of the coverage up to $100$ time units for various values of the size ratio, namely, $1$, $1.1$, $2$, $4$, and $10$.
    $\mbox{b)}$ Plot showing the jamming coverage dependence on the size ratio.
   \label{jammingcoverage}}
  \end{figure}

  We start by presenting our results with a study of the coverage for the particular case of equal fluxes of incoming segments.
  In this respect, our study represents the particular case of $q=1/2$ of reference \cite{Hassan02}, but with a wider range of size ratios.
  In fact, results presented in \cite{Hassan02} are valid for values of $R<2$.
  Our study also includes quantities, such as the dispersion, the fraction of empty spaces, etc., which are not possible to be computed by their method.
  In Fig.~\ref{jammingcoverage}(a), we show the time dependence of the coverage, $\theta(t)$, for some of the values we simulated of the size ratio, $R$.
  However, we performed simulations for a wider range of values of the size ratio, namely for values of $1$, $1.05$, $1.1$, $1.2$, $1.25$, $1.3$, $1.4$, $1.5$, $1.6$, $1.75$, $2$, $2.1$, $2.25$, $2.5$, $2.75$, $3$, $3.25$, $3.5$, $3.75$, $4$, $4.5$, $5$, $5.5$, $6$, $7$, $8$, $10$, $12$, $14$, $16$, $18$, and $20$.
  Notice that our model for a size ratio of one boils down to deposition of segments of unit size on the line.
  The value of $74.759 58 \% \pm 0.0067 \%$ for the jamming coverage is quite close to $74.759 792 02 \%$, first obtained by R\'enyi and subsequently calculated with larger precision by Blaisdell and Solomon \cite{Renyi58,Hassan02,Bonnier01}.
  \footnote{The quoted result represents the first ten of $14$ decimal figures reckoned by Blaisdell and Solomon \cite{Blaisdell70}.}

  \begin{figure}
   \includegraphics[width=8.5cm]{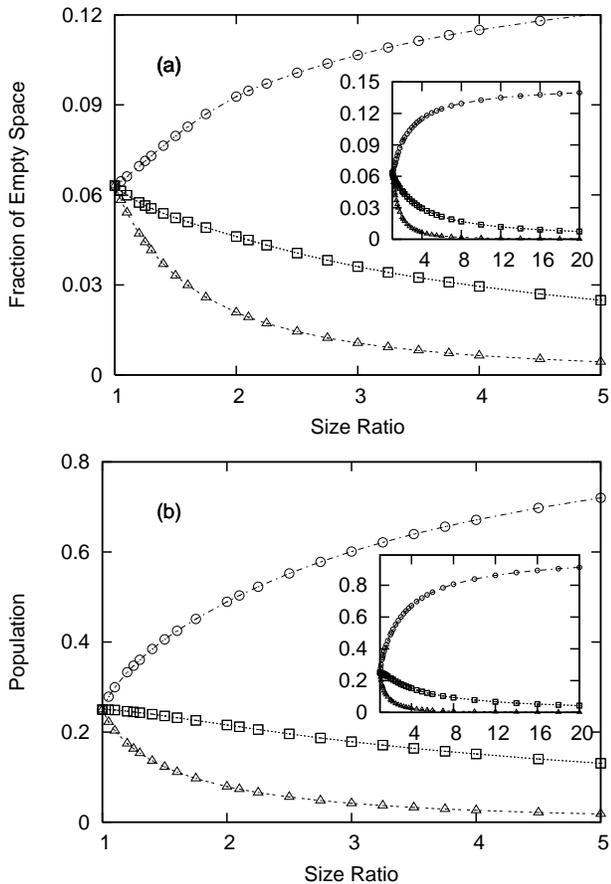}
   \caption{
    In both plots, one has  the $AA$-, $AB$-, and $BB$-gap types represented by circles, squares, and triangles, respectively.
     a) Fraction of available empty space, at the jamming state.
     b) Normalized population of gaps at the jamming state.
    In the insets the size ratio varies between $1$ and $20$.
    Please, refer to the text for further details.
    \label{empty}}
  \end{figure}

  The jamming coverage increases monotonically as a function of the size ratio, as shown in Fig.~\ref{jammingcoverage}(b) and Table~\ref{table:jc}.
  This result is not entirely surprising, since for equal relative fluxes of incoming segments and for large, asymptotic values of the size ratio, a stretch of the line is either fully covered by a large segment or paved by the small, unit size, segments with an upper limit of the coverage given by the above R\'enyi value.
  As both segment sizes can attempt deposition with equal probability, we get $1/2(1+0.7476)\equiv87.38 \%$.
  However, this upper limit is not attained as one increases the value of the size ratio: the limiting value of the coverage around $84.5 \%$, obtained for a size ratio of $20$, remains lower as one can observe in Fig.~\ref{jammingcoverage}(b).

  In order to better analyze the structure of the jamming state, we now proceed to characterize the fraction of empty space distributed in terms of the different pairs of segments as we vary the size ratio.
  In Fig.~\ref{empty}(a), we plot the fraction of the substrate left empty at the jamming state as a function of the size ratio for each pair of consecutive segments, namely of type $AA$, $AB$, and $BB$.
  In Fig.~\ref{empty}(b), we also present the corresponding normalized population of the gap types.
  Consequently, in the latter case, the sum of the contributions of the four gap types adds up to the unit for every value of the size ratio.
  We observe that the fraction of empty space due to the $AA\mbox{-gaps}$ increases monotonically with the size ratio.
  This can be understood by considering the significant drop in the population of $BB\mbox{-gaps}$ at the jamming state as the size ratio increases, since these gaps must be smaller than unit.
  The probability of having two consecutive $B$-segments adsorbed on the line rapidly decreases with the size ratio.
  Not only does the adsorption of both large segments imply a clean substrate, for example by the absence of smaller segments previously adsorbed on the line, but also both large segments must adsorb at a distance smaller than unit to prevent an $A\mbox{-segment}$ to fit in.
  The latter situation becomes less and less probable as the size ratio increases, and, to reach the jamming state, the leftover empty space between these large segments must be filled with segments of unit size, thus, contributing to a higher fraction of empty space associated with $AA\mbox{-gaps}$ and, consequently, to a higher value of the coverage.
  The slower decrease of the $AB\mbox{-gap}$ population and of the corresponding coverage indicates a jamming state constituted by alternating streaks of $A\mbox{-segments}$ separated by a single $B\mbox{-segment}$.
  This argument is further substantiated by the strong decay of the fraction of empty space associated with the $BB\mbox{-gaps}$ for increasing values of the size ratio.
  This is also corroborated by the smoother decay of both the fraction of empty space and of the population of the $AB\mbox{-gap}$ type, due to the breakdown of the $BB\mbox{-gap}$ type as the size ratio increases.
  Finally, it can be seen in Fig.~\ref{empty}(b) that as the size ratio diverges, the relative population of $AB\mbox{-}$ and $BB\mbox{-gap}$ types as compared to the corresponding population of the $AA\mbox{-gaps}$ becomes less and less relevant, therefore leaving, in this limit, long streaks of $A\mbox{-segments}$.

  \begin{figure}
   \includegraphics[width=8.5cm]{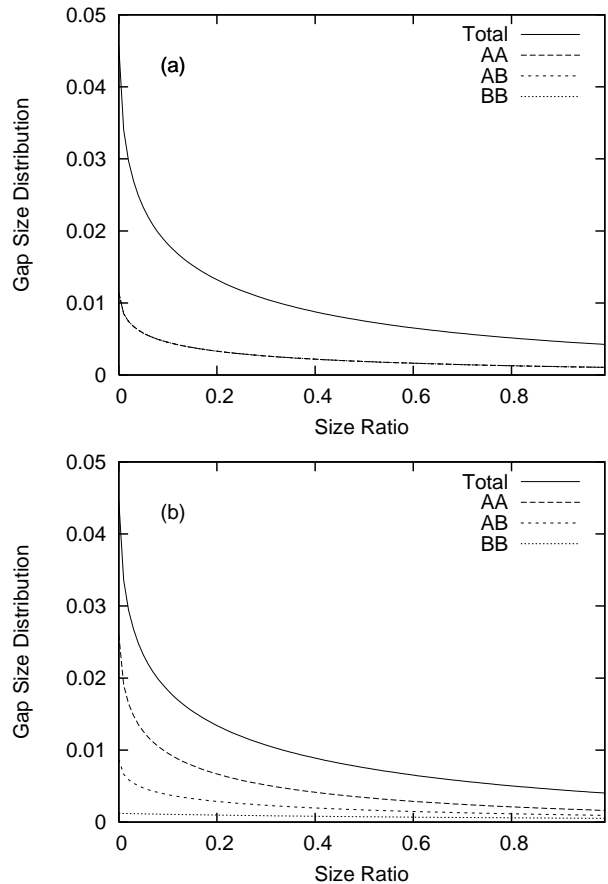}
   \caption{
    Gap-size distribution functions at the jamming state:
     a) For a size ratio of one, the solid curve represents $P_\emptyset(x)$, while the remaining $AA\mbox{-}$, $AB\mbox{-}$, and $BB\mbox{-gap}$ types are identical, and therefore fall onto a single curve.
     b) For a size ratio of $2.0$, one observes the splitting of the various gap types.
     Please, refer to the text for further details.
    \label{distribution}}
  \end{figure}

  \begin{table}
    \caption{
      The table summarizes the estimated values of $R$, where the first four cumulants take maximum and minimum values, all measured with an error of $\pm 0.005$.
      The corresponding values of the cumulants are also included.
      \label{table:minmax}
    }
    \begin{ruledtabular}
      \begin{tabular}{c c c c c c}
        &gap types&\multicolumn{2}{c}{minimum}&\multicolumn{2}{c}{maximum}\\
        &&$R$&Value&$R$&Value\\
        \hline\\
        Distance&&&\\
        &AA&1.31&0.2924895&&\\
        &AB&1.22&0.3291302&&\\
        &BB&\multicolumn{2}{c}{$^($\footnote{The $BB\mbox{-gap}$ is strictly monotonic.\label{table:minmax:first}}$^)$}&\multicolumn{2}{c}{$^{(\mbox{\tiny\ref{table:minmax:first}})}$}\\
        Dispersion&&&\\
        &AA&1.55&0.2681662&1.02&0.2825507\\
        &AB&1.55&0.2788692&1.05&0.2825593\\
        &BB&1.01&0.2815730&-&-\\
        Skewness&&&\\
        &AA&-&-&1.34&0.9227950\\
        &AB&-&-&1.29&0.7253401\\
        &BB&\multicolumn{2}{c}{$^{(\mbox{\tiny\ref{table:minmax:first}})}$}&\multicolumn{2}{c}{$^{(\mbox{\tiny\ref{table:minmax:first}})}$}\\
        Kurtosis&&&\\
        &AA&-&-&1.40&$-$0.2290402\\
        &AB&-&-&1.36&$-$0.6281340\\
        &BB&\multicolumn{2}{c}{$^{(\mbox{\tiny\ref{table:minmax:first}})}$}&\multicolumn{2}{c}{$^{(\mbox{\tiny\ref{table:minmax:first}})}$}\\
      \end{tabular}
    \end{ruledtabular}
  \end{table}

  In Fig.~\ref{distribution}(a), we show the distribution functions, $P_\emptyset(x)$, $P_{AA}(x)$, $P_{AB}(x)$, and $P_{BB}(x)$, at the jamming state, for a size ratio of one.
  Of course, at this size ratio ($R=1$), all gap types are equal as there is a single segment size.
  Therefore, one expects, as shown in Fig.~\ref{distribution}(a), to observe the collapse of the $P_{AA}(x)$, $P_{AB}(x)$, and $P_{BB}(x)$, onto a single curve.
  The $P_\emptyset(x)$ distribution function also satisfies the relation $P_\emptyset(x)/4=P_{AA}(x)=P_{AB}(x)=P_{BB}(x)$ in agreement with equation (\ref{discrimination}) for a size ratio of one.
  In the case of part b) of the same figure, we merely changed the value of the size ratio to $2.0$, thus breaking the symmetries between the various gap types.
  Once again, adding the various gap distribution functions, for every value of x, accordingly to equation (\ref{discrimination}), actually reproduces $P_\emptyset(x)$.
  As compared to part a), the $AA\mbox{-gap}$ type increases its dominance, as pointed out above, while the $AB\mbox{-gap}$ slightly lowers its contribution.
  The $BB\mbox{-gap}$ significantly drops its influence, because it becomes less probable to have consecutive deposition of $B\mbox{-segments}$ with gap lengths smaller than unit.

  \begin{figure}
   \includegraphics[width=8.5cm]{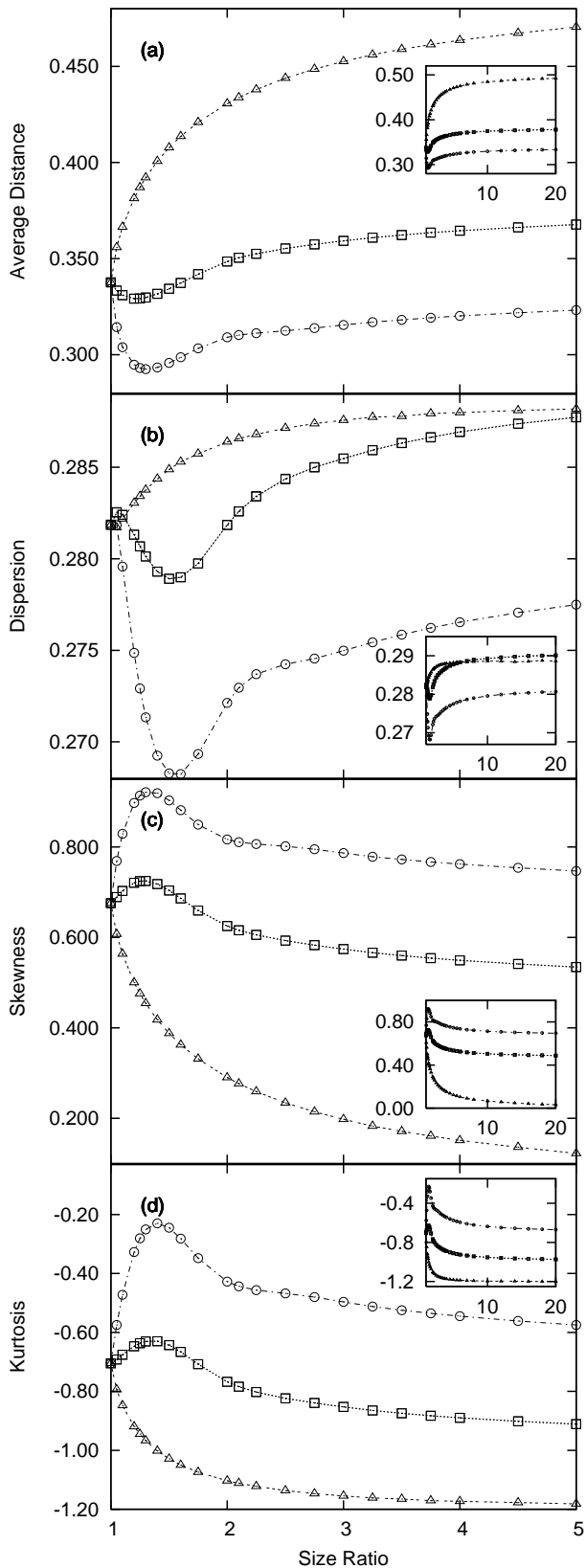}
   \caption{
     Plots involving cumulants up to the fourth order of the gap-size distribution functions for each gap type as a function of the size ratio.
     We use the same legend as in Fig.~\ref{empty}.
     a) Average distance between pairs of segments.
     b) Dispersion of the distance between pairs of segments.
     c) Skewness.
     d) Kurtosis.
   \label{cum}}
  \end{figure}

  Now, we proceed to the analysis of quantities involving cumulants up to the fourth order, more specifically, the mean distance, dispersion, skewness, and kurtosis, as defined in equations (\ref{cumulant_k1}$\mbox{-}$\ref{kurtosis}).
  For each of the latter quantities, we studied their dependence on the size ratio at the jamming coverage limit as shown in Fig.~\ref{cum}.
  Both the $AA\mbox{-}$ and $AB\mbox{-gap}$ types have rich non-monotonic behavior, for the mean gap size, dispersion, skewness and kurtosis.
  For example, the minimum values of the gap size for the $AA\mbox{-}$ and $AB\mbox{-gaps}$ are not coincidental, as the minimum of the AA-gap occurs at a slightly higher value of the size ratio.
  The behavior of the dispersion shows the presence of both a minimum and maximum values.
  However, the opposite happens for the maximum value: the value of the size ratio at which it occurs is slightly lower for the $AA\mbox{-gap}$ type.
  The minimum value occurs at the same value of $R$ for these two gap types, within the error bounds.
  The skewness shows a maximum occurring at the different $R$ values, namely, $1.34$ and $1.29$ for the $AA\mbox{-}$ and $AB\mbox{-gap}$ types, respectively.
  The kurtosis reveals a single maximum for these two gap types, once again non-coincidental in their size ratio value, with the maximum of the size ratio of the $AA$ gap slightly above, than the corresponding one found for the $AB$ gap.
  All these values of $R$ are summarized in Table~\ref{table:minmax}.
  Regarding the $BB\mbox{-gap}$ type, one finds a less rich behavior as compared with the other gap types.
  For this gap type the mean gap size strictly increases monotonically with $R$, while it decreases strictly monotonically for both the skewness and kurtosis.
  The exception to this behavior for the $BB$ gap regards the dispersion, where it shows a minimum for $R=1.01$ as shown in Table~\ref{table:minmax} and Fig.~\ref{cum}(b).
  Finally, the dispersion of both the $BB\mbox{-}$ and $AB\mbox{-gap}$ types intersect at $R=6$, as shown in the inset of Fig.~\ref{cum}(b).

  If one were to solely deposit large segments, then all gap sizes smaller than $R$ would be present.
  Since we are concomitantly depositing unit segments, all gap sizes larger than unit must disappear at the jamming state.
  As soon as the typical size of gaps drops below R, the large segments stop adsorbing on the line.
  The smaller, unit segments, contrary to  larger segments, have to fit into all available space, i.e., in all gaps larger than unit.
  For size ratios $R<1.55$ the population of $BB\mbox{-gaps}$ remains significant, as compared to the $AA\mbox{-gap}$ one, with the consequent formation of $B\mbox{-segments}$ gap sizes close to unit.
  Some of these gaps are really close to unit, but strictly larger than unit, in size, and we call these events {\it snug fits}.
  As $R$ increases the number of $BB\mbox{-gaps}$ decreases, to form $AA\mbox{-}$ and $AB\mbox{-gaps}$, therefore, increasing the probability of {\it snug fit} events.
  The probability that for large values of $R$ $BB\mbox{-gaps}$ form a {\it snug fit} also decreases.
  Consequently, the importance of {\it snug fits} decreases for large values of $R$ in agreement with Fig.~\ref{cum}, where one observes monotonic behavior of the various cumulants in this regime.
  However, the large number of these events for values of $R\lesssim1.55$ accounts for such rich behavior of the cumulants up to the fourth order.
  For example, the existence of minimum values of the $AA\mbox{-}$ and $AB\mbox{-gaps}$ sizes can now be understood as events from late stage kinetics close to the jamming state.
  The same argument also applies to the minimum values of the dispersion, since this events tend to lower the uncertainties of adsorbed segments.
  More suble, it is the presence of maximum values for the both the $AA\mbox{-}$ and $AB\mbox{-gaps}$ and a minimum of the $BB\mbox{-gaps}$ for values of $R<1.05$.
  Since the difference in size of both segments is small, both segment sizes compete until coverage values become close to the jamming state, thus leading to {\it snug fit} events of the $BB\mbox{-gap}$.
  This competition leads to lessen the uncertainty in the $BB\mbox{-gaps}$, but it increases the uncertainties of the $AA\mbox{-}$ and $AB\mbox{-gap}$ types.
  For values of $R\ge1.55$, the significant drop in the population of $BB\mbox{-gaps}$ can be understood of the early onset of a mean gap size smaller than $R$ at low coverage values, which effectively blocks the adsorption of large segments.
  Consequently, one expects a flatter distribution function of the $BB\mbox{-gap}$ type with increasing values of the size ratio, i.e., the net effect of increasing the size ratio is to diminish the asymmetry of the $BB$-gap distribution function.
  The monotonic behavior of the $BB\mbox{-gap}$ for large size ratios ($R > 1.55$) stems from the absence of a competing mechanism.
  Finally, the non-monotonic behavior of the remaining gap types is due to \textit{snug fit} events, which the latter tend to askew the corresponding gap-size distribution functions by favoring small gap sizes.

  The jamming state for large size ratios, follows, therefore, the picture of alternating streaks of $A\mbox{-segments}$ interfaced with a single $B\mbox{-segment}$ for large values of the size ratio.
  The presence of such streaks of $A\mbox{-segments}$ interfaced with a single $B\mbox{-segment}$, prevents the $AB\mbox{-gap}$ population to decrease slower than that of the $BB\mbox{-gap}$ one.
  The \textit{snug fits} events also tend to favor small gap sizes, and this effect makes the distribution function less \textit{flat} and more askewed, thus increasing the value of the kurtosis and skewness of both the $AA\mbox{-}$ and $AB\mbox{-gaps}$. (Fig.~\ref{cum}(c-d) and Fig.~\ref{distribution})

 \section{Conclusion\label{conclusion}}
  We analyzed in detail the jamming structure of a model of random sequential adsorption on the line with two-segment sizes depositing with equal fluxes.
  The structure of the jamming state is determined by the population of the various gap types, namely, the $AA\mbox{-}$, $AB\mbox{-}$, and $BB\mbox{-gap}$ types.
  For large values of the size ratio, i.e., for size ratios greater than $1.55$, \textit{snug fits} become less significant, and the average distance, dispersion, skewness, and kurtosis monotonically approach their asymptotic values.
  For values of the size ratio below two, the behavior of the above quantities for the $AA\mbox{-}$, and $AB\mbox{-}$ types are non-monotonic due to the presence of \textit{snug fit} events.
  The jamming state at values of $R$ above $1.55$ is characterized by streaks of $A\mbox{-}$ interrupted by a single $B\mbox{-segment}$.
  At sizes ratios smaller than $1.55$ a rich, non-monotonic behavior of the above quantities plays develops due to the interplay provided by {\it snug fits}.
  From an experimental point of view, the study provides to some extend, among others, the insight on how to tune up the mean interparticle distance or how to minimize the flutuactions of the interparticle distance around the mean value.

 \begin{acknowledgments}
  This research has been funded by two Funda\c c\~ao para a Ci\^encia e a Tecnologia research grants: Computational Nanophysics (under contract POCTI/CTM/41574/2001) and SeARCH (Services and Advanced Research Computing with HTC/HPC clusters) (under contract CONC-REEQ/443/2001).
  One of us (NA) thanks Funda\c c\~ao para a Ci\^encia e a Tecnologia for a Ph.D grant (SFRH/BD/17467/2004).
  We also want to thank Professor Vladimir Privman for useful comments.
 \end{acknowledgments}

 \bibliography{text}

\begin{thebibliography}{38}
\expandafter\ifx\csname natexlab\endcsname\relax\def\natexlab#1{#1}\fi
\expandafter\ifx\csname bibnamefont\endcsname\relax
  \def\bibnamefont#1{#1}\fi
\expandafter\ifx\csname bibfnamefont\endcsname\relax
  \def\bibfnamefont#1{#1}\fi
\expandafter\ifx\csname citenamefont\endcsname\relax
  \def\citenamefont#1{#1}\fi
\expandafter\ifx\csname url\endcsname\relax
  \def\url#1{\texttt{#1}}\fi
\expandafter\ifx\csname urlprefix\endcsname\relax\def\urlprefix{URL }\fi
\providecommand{\bibinfo}[2]{#2}
\providecommand{\eprint}[2][]{\url{#2}}

\bibitem[{\citenamefont{Flory}(1939)}]{Flory39}
\bibinfo{author}{\bibfnamefont{P.}~\bibnamefont{Flory}}, \bibinfo{journal}{J.
  Am. Chem. Soc.} \textbf{\bibinfo{volume}{61}}, \bibinfo{pages}{1518}
  (\bibinfo{year}{1939}).

\bibitem[{\citenamefont{R\'enyi}(1958)}]{Renyi58}
\bibinfo{author}{\bibfnamefont{A.}~\bibnamefont{R\'enyi}},
  \bibinfo{journal}{Publ. Math. Inst. Hung. Acad. Sci.}
  \textbf{\bibinfo{volume}{3}}, \bibinfo{pages}{109} (\bibinfo{year}{1958}).

\bibitem[{\citenamefont{R\'enyi}(1963)}]{Renyi63}
\bibinfo{author}{\bibfnamefont{A.}~\bibnamefont{R\'enyi}},
  \bibinfo{journal}{Sel. Trans. Math. Stat. Prob.}
  \textbf{\bibinfo{volume}{4}}, \bibinfo{pages}{203} (\bibinfo{year}{1963}).

\bibitem[{\citenamefont{Gonz\'alez et~al.}(1974)\citenamefont{Gonz\'alez,
  Hemmer, and H{\o}ye}}]{Gonzalez74}
\bibinfo{author}{\bibfnamefont{J.}~\bibnamefont{Gonz\'alez}},
  \bibinfo{author}{\bibfnamefont{P.}~\bibnamefont{Hemmer}}, \bibnamefont{and}
  \bibinfo{author}{\bibfnamefont{J.}~\bibnamefont{H{\o}ye}},
  \bibinfo{journal}{Chem. Phys.} \textbf{\bibinfo{volume}{3}},
  \bibinfo{pages}{288} (\bibinfo{year}{1974}).

\bibitem[{\citenamefont{Evans}(1993)}]{Evans93}
\bibinfo{author}{\bibfnamefont{J.}~\bibnamefont{Evans}}, \bibinfo{journal}{Rev.
  Mod. Phys.} \textbf{\bibinfo{volume}{65}}, \bibinfo{pages}{1281}
  (\bibinfo{year}{1993}).

\bibitem[{\citenamefont{Privman}(1994)}]{Privman94}
\bibinfo{author}{\bibfnamefont{V.}~\bibnamefont{Privman}},
  \bibinfo{journal}{Trends in Stat. Phys.} \textbf{\bibinfo{volume}{1}},
  \bibinfo{pages}{89} (\bibinfo{year}{1994}).

\bibitem[{\citenamefont{Privman(Ed.)}(1997)}]{Privman97}
\bibinfo{author}{\bibfnamefont{V.}~\bibnamefont{Privman(Ed.)}},
  \emph{\bibinfo{title}{Nonequilibrium Statistical Mechanics}}
  (\bibinfo{publisher}{Cambridge University Press},
  \bibinfo{address}{Cambridge, United Kingdom}, \bibinfo{year}{1997}).

\bibitem[{\citenamefont{Marro and Dickman}(1999)}]{Marro99}
\bibinfo{author}{\bibfnamefont{J.}~\bibnamefont{Marro}} \bibnamefont{and}
  \bibinfo{author}{\bibfnamefont{R.}~\bibnamefont{Dickman}},
  \emph{\bibinfo{title}{Nonequilibrium Phase Transitions in Lattice Models}}
  (\bibinfo{publisher}{Cambridge University Press},
  \bibinfo{address}{Cambridge, United Kingdom}, \bibinfo{year}{1999}).

\bibitem[{\citenamefont{Privman}(2000{\natexlab{a}})}]{Privman00a}
\bibinfo{author}{\bibfnamefont{V.}~\bibnamefont{Privman}}, \bibinfo{journal}{J.
  Adhesion} \textbf{\bibinfo{volume}{74}}, \bibinfo{pages}{42}
  (\bibinfo{year}{2000}{\natexlab{a}}).

\bibitem[{\citenamefont{Privman}(2000{\natexlab{b}})}]{Privman00b}
\bibinfo{author}{\bibfnamefont{V.}~\bibnamefont{Privman}},
  \bibinfo{journal}{Coll. and Surf. A} \textbf{\bibinfo{volume}{165}},
  \bibinfo{pages}{231} (\bibinfo{year}{2000}{\natexlab{b}}).

\bibitem[{\citenamefont{Onoda and Liniger}(1986)}]{Onoda86}
\bibinfo{author}{\bibfnamefont{G.}~\bibnamefont{Onoda}} \bibnamefont{and}
  \bibinfo{author}{\bibfnamefont{E.}~\bibnamefont{Liniger}},
  \bibinfo{journal}{Phys. Rev. A} \textbf{\bibinfo{volume}{33}},
  \bibinfo{pages}{715} (\bibinfo{year}{1986}).

\bibitem[{\citenamefont{Chen et~al.}(2002)\citenamefont{Chen, Klemic, and
  Elimelech}}]{Chen02}
\bibinfo{author}{\bibfnamefont{J.}~\bibnamefont{Chen}},
  \bibinfo{author}{\bibfnamefont{J.}~\bibnamefont{Klemic}}, \bibnamefont{and}
  \bibinfo{author}{\bibfnamefont{M.}~\bibnamefont{Elimelech}},
  \bibinfo{journal}{Nano Lett.} \textbf{\bibinfo{volume}{2}},
  \bibinfo{pages}{393} (\bibinfo{year}{2002}).

\bibitem[{\citenamefont{Cadilhe and Privman}(2004)}]{Cadilhe04}
\bibinfo{author}{\bibfnamefont{A.}~\bibnamefont{Cadilhe}} \bibnamefont{and}
  \bibinfo{author}{\bibfnamefont{V.}~\bibnamefont{Privman}},
  \bibinfo{journal}{Mod. Phys. Lett. B} \textbf{\bibinfo{volume}{18}},
  \bibinfo{pages}{207} (\bibinfo{year}{2004}).

\bibitem[{\citenamefont{Ara\'ujo and Cadilhe}()}]{Araujo04a}
\bibinfo{author}{\bibfnamefont{N.}~\bibnamefont{Ara\'ujo}} \bibnamefont{and}
  \bibinfo{author}{\bibfnamefont{A.}~\bibnamefont{Cadilhe}}, \bibinfo{note}{to
  be published elsewhere}.

\bibitem[{\citenamefont{Tassel et~al.}(2000)\citenamefont{Tassel, Viot, Tarjus,
  Ramsden, and Talbot}}]{Tassel00}
\bibinfo{author}{\bibfnamefont{P.~V.} \bibnamefont{Tassel}},
  \bibinfo{author}{\bibfnamefont{P.}~\bibnamefont{Viot}},
  \bibinfo{author}{\bibfnamefont{G.}~\bibnamefont{Tarjus}},
  \bibinfo{author}{\bibfnamefont{J.}~\bibnamefont{Ramsden}}, \bibnamefont{and}
  \bibinfo{author}{\bibfnamefont{J.}~\bibnamefont{Talbot}},
  \bibinfo{journal}{J. Chem. Phys.} \textbf{\bibinfo{volume}{112}},
  \bibinfo{pages}{1483} (\bibinfo{year}{2000}).

\bibitem[{\citenamefont{Krapivsky and Ben-Naim}(1994)}]{Krapivsky94}
\bibinfo{author}{\bibfnamefont{P.}~\bibnamefont{Krapivsky}} \bibnamefont{and}
  \bibinfo{author}{\bibfnamefont{E.}~\bibnamefont{Ben-Naim}},
  \bibinfo{journal}{J. Chem. Phys.} \textbf{\bibinfo{volume}{100}},
  \bibinfo{pages}{6778} (\bibinfo{year}{1994}).

\bibitem[{\citenamefont{Bonnier}(1997)}]{Bonnier97}
\bibinfo{author}{\bibfnamefont{B.}~\bibnamefont{Bonnier}},
  \bibinfo{journal}{Phys. Rev. E} \textbf{\bibinfo{volume}{56}},
  \bibinfo{pages}{7304} (\bibinfo{year}{1997}).

\bibitem[{\citenamefont{Nielaba}()}]{Nielaba97}
\bibinfo{author}{\bibfnamefont{P.}~\bibnamefont{Nielaba}}, \bibinfo{note}{in
  ref. \cite{Privman97}, p. 229}.

\bibitem[{\citenamefont{Nielaba and Privman}(1995)}]{Nielaba95}
\bibinfo{author}{\bibfnamefont{P.}~\bibnamefont{Nielaba}} \bibnamefont{and}
  \bibinfo{author}{\bibfnamefont{V.}~\bibnamefont{Privman}},
  \bibinfo{journal}{Phys. Rev. E} \textbf{\bibinfo{volume}{51}},
  \bibinfo{pages}{2022} (\bibinfo{year}{1995}).

\bibitem[{\citenamefont{Privman and Wang}(1992)}]{Privman92}
\bibinfo{author}{\bibfnamefont{V.}~\bibnamefont{Privman}} \bibnamefont{and}
  \bibinfo{author}{\bibfnamefont{J.-S.} \bibnamefont{Wang}},
  \bibinfo{journal}{Phys. Rev. A} \textbf{\bibinfo{volume}{45}},
  \bibinfo{pages}{R2155} (\bibinfo{year}{1992}).

\bibitem[{\citenamefont{Fan and Percus}(1991{\natexlab{a}})}]{Fan91a}
\bibinfo{author}{\bibfnamefont{Y.}~\bibnamefont{Fan}} \bibnamefont{and}
  \bibinfo{author}{\bibfnamefont{J.}~\bibnamefont{Percus}},
  \bibinfo{journal}{Phys. Rev. Lett.} \textbf{\bibinfo{volume}{67}},
  \bibinfo{pages}{1677} (\bibinfo{year}{1991}{\natexlab{a}}).

\bibitem[{\citenamefont{Fan and Percus}(1991{\natexlab{b}})}]{Fan91b}
\bibinfo{author}{\bibfnamefont{Y.}~\bibnamefont{Fan}} \bibnamefont{and}
  \bibinfo{author}{\bibfnamefont{J.}~\bibnamefont{Percus}},
  \bibinfo{journal}{Phys. Rev. A} \textbf{\bibinfo{volume}{44}},
  \bibinfo{pages}{5099} (\bibinfo{year}{1991}{\natexlab{b}}).

\bibitem[{\citenamefont{Bartelt}(1991)}]{Bartelt91}
\bibinfo{author}{\bibfnamefont{M.}~\bibnamefont{Bartelt}},
  \bibinfo{journal}{Phys. Rev. A} \textbf{\bibinfo{volume}{43}},
  \bibinfo{pages}{3149} (\bibinfo{year}{1991}).

\bibitem[{\citenamefont{Bonnier et~al.}(1994)\citenamefont{Bonnier,
  Hontebeyrie, Leroyer, Meyers, and Pommiers}}]{Bonnier94}
\bibinfo{author}{\bibfnamefont{B.}~\bibnamefont{Bonnier}},
  \bibinfo{author}{\bibfnamefont{M.}~\bibnamefont{Hontebeyrie}},
  \bibinfo{author}{\bibfnamefont{Y.}~\bibnamefont{Leroyer}},
  \bibinfo{author}{\bibfnamefont{C.}~\bibnamefont{Meyers}}, \bibnamefont{and}
  \bibinfo{author}{\bibfnamefont{E.}~\bibnamefont{Pommiers}},
  \bibinfo{journal}{Phys. Rev. E} \textbf{\bibinfo{volume}{49}},
  \bibinfo{pages}{305} (\bibinfo{year}{1994}).

\bibitem[{\citenamefont{Evans}()}]{Evans97}
\bibinfo{author}{\bibfnamefont{J.}~\bibnamefont{Evans}}, \bibinfo{note}{in ref.
  \cite{Privman97}, p.205}.

\bibitem[{\citenamefont{Blaisdell and Solomon}(1970)}]{Blaisdell70}
\bibinfo{author}{\bibfnamefont{B.}~\bibnamefont{Blaisdell}} \bibnamefont{and}
  \bibinfo{author}{\bibfnamefont{H.}~\bibnamefont{Solomon}},
  \bibinfo{journal}{J. Appl. Prob.} \textbf{\bibinfo{volume}{7}},
  \bibinfo{pages}{667} (\bibinfo{year}{1970}).

\bibitem[{\citenamefont{Privman et~al.}(1991)\citenamefont{Privman, Wang, and
  Nielaba}}]{Privman91}
\bibinfo{author}{\bibfnamefont{V.}~\bibnamefont{Privman}},
  \bibinfo{author}{\bibfnamefont{J.-S.} \bibnamefont{Wang}}, \bibnamefont{and}
  \bibinfo{author}{\bibfnamefont{P.}~\bibnamefont{Nielaba}},
  \bibinfo{journal}{Phys. Rev. B} \textbf{\bibinfo{volume}{43}},
  \bibinfo{pages}{3366} (\bibinfo{year}{1991}).

\bibitem[{\citenamefont{Bartelt and Evans}(1994)}]{Bartelt94}
\bibinfo{author}{\bibfnamefont{M.}~\bibnamefont{Bartelt}} \bibnamefont{and}
  \bibinfo{author}{\bibfnamefont{J.}~\bibnamefont{Evans}}, \bibinfo{journal}{J.
  Stat. Phys.} \textbf{\bibinfo{volume}{76}}, \bibinfo{pages}{867}
  (\bibinfo{year}{1994}).

\bibitem[{\citenamefont{Dickman et~al.}(1991)\citenamefont{Dickman, Wang, and
  Jensen}}]{Dickman91}
\bibinfo{author}{\bibfnamefont{R.}~\bibnamefont{Dickman}},
  \bibinfo{author}{\bibfnamefont{J.-S.} \bibnamefont{Wang}}, \bibnamefont{and}
  \bibinfo{author}{\bibfnamefont{I.}~\bibnamefont{Jensen}},
  \bibinfo{journal}{J. Chem. Phys.} \textbf{\bibinfo{volume}{94}},
  \bibinfo{pages}{8252} (\bibinfo{year}{1991}).

\bibitem[{\citenamefont{Evans et~al.}(1984)\citenamefont{Evans, Burgess, and
  Hoffman}}]{Evans84a}
\bibinfo{author}{\bibfnamefont{J.}~\bibnamefont{Evans}},
  \bibinfo{author}{\bibfnamefont{D.}~\bibnamefont{Burgess}}, \bibnamefont{and}
  \bibinfo{author}{\bibfnamefont{D.}~\bibnamefont{Hoffman}},
  \bibinfo{journal}{J. Math Phys.} \textbf{\bibinfo{volume}{25}},
  \bibinfo{pages}{3051} (\bibinfo{year}{1984}).

\bibitem[{\citenamefont{Privman}(1993)}]{Privman93}
\bibinfo{author}{\bibfnamefont{V.}~\bibnamefont{Privman}},
  \bibinfo{journal}{Europhys. Lett.} \textbf{\bibinfo{volume}{23}},
  \bibinfo{pages}{341} (\bibinfo{year}{1993}).

\bibitem[{\citenamefont{Bonnier}(2001)}]{Bonnier01}
\bibinfo{author}{\bibfnamefont{B.}~\bibnamefont{Bonnier}},
  \bibinfo{journal}{Phys. Rev. E} \textbf{\bibinfo{volume}{64}},
  \bibinfo{pages}{066111} (\bibinfo{year}{2001}).

\bibitem[{\citenamefont{Hassan and Kurths}(2001)}]{Hassan01}
\bibinfo{author}{\bibfnamefont{M.~K.} \bibnamefont{Hassan}} \bibnamefont{and}
  \bibinfo{author}{\bibfnamefont{J.}~\bibnamefont{Kurths}},
  \bibinfo{journal}{J. Phys. A} \textbf{\bibinfo{volume}{34}},
  \bibinfo{pages}{7517} (\bibinfo{year}{2001}).

\bibitem[{\citenamefont{Hassan et~al.}(2002)\citenamefont{Hassan, Schmidt,
  Blasius, and Kurths}}]{Hassan02}
\bibinfo{author}{\bibfnamefont{M.}~\bibnamefont{Hassan}},
  \bibinfo{author}{\bibfnamefont{J.}~\bibnamefont{Schmidt}},
  \bibinfo{author}{\bibfnamefont{B.}~\bibnamefont{Blasius}}, \bibnamefont{and}
  \bibinfo{author}{\bibfnamefont{J.}~\bibnamefont{Kurths}},
  \bibinfo{journal}{Phys. Rev. E} \textbf{\bibinfo{volume}{65}},
  \bibinfo{pages}{045103} (\bibinfo{year}{2002}).

\bibitem[{\citenamefont{Brilliantov et~al.}(1996)\citenamefont{Brilliantov,
  Andrienko, Krapivsky, and Kurths}}]{Brilliantov96}
\bibinfo{author}{\bibfnamefont{N.}~\bibnamefont{Brilliantov}},
  \bibinfo{author}{\bibfnamefont{Y.}~\bibnamefont{Andrienko}},
  \bibinfo{author}{\bibfnamefont{P.}~\bibnamefont{Krapivsky}},
  \bibnamefont{and} \bibinfo{author}{\bibfnamefont{J.}~\bibnamefont{Kurths}},
  \bibinfo{journal}{Phys. Rev. Lett.} \textbf{\bibinfo{volume}{76}},
  \bibinfo{pages}{4058} (\bibinfo{year}{1996}).

\bibitem[{\citenamefont{Evans}(1984)}]{Evans84b}
\bibinfo{author}{\bibfnamefont{J.}~\bibnamefont{Evans}}, \bibinfo{journal}{J.
  Math. Phys.} \textbf{\bibinfo{volume}{25}}, \bibinfo{pages}{2527}
  (\bibinfo{year}{1984}).

\bibitem[{\citenamefont{Swendsen}(1981)}]{Swendsen81}
\bibinfo{author}{\bibfnamefont{R.}~\bibnamefont{Swendsen}},
  \bibinfo{journal}{Phys. Rev. A} \textbf{\bibinfo{volume}{24}},
  \bibinfo{pages}{504} (\bibinfo{year}{1981}).

\bibitem[{\citenamefont{van Kampen}(1992)}]{Kampen92}
\bibinfo{author}{\bibfnamefont{N.}~\bibnamefont{van Kampen}},
  \emph{\bibinfo{title}{Stochastic Processes in Physics and Chemistry}}
  (\bibinfo{publisher}{Elsevier Science Publishers},
  \bibinfo{address}{Amesterdam, The Netherlands}, \bibinfo{year}{1992}).

\end{thebibliography}
\end{document}